\documentclass{article}
\usepackage[left=3.7cm,right=3.7cm,top=3.5cm,bottom=3.4cm]{geometry}
\usepackage{graphicx} 
\usepackage{caption}
\usepackage{chemformula}
\usepackage{cite}
\usepackage{mathtools}
\usepackage{adjustbox}
\usepackage{amsmath}
\usepackage[dvipsnames]{xcolor}
\usepackage{setspace}
\usepackage{geometry}
\usepackage{gensymb}
\usepackage{siunitx}
\usepackage{upgreek}

\begin{document}

\begin{center}

{\Large\textbf{Suppression of stripe-ordered structural phases in monolayer \ch{IrTe2} by a gold substrate}} \\[8pt]
Kati Asikainen\textsuperscript{\emph{1,2}}, Fr$\mathrm{\acute{e}}$d$\mathrm{\acute{e}}$ric Chassot\textsuperscript{\emph{2}}, Baptiste Hildebrand\textsuperscript{\emph{2}}, Aymen Mahmoudi\textsuperscript{\emph{2}}, Joel Morf\textsuperscript{\emph{2}}, Mahault Berset\textsuperscript{\emph{2}}, Pascal Turban\textsuperscript{\emph{3}}, Matti Alatalo\textsuperscript{\emph{1}}, S. Assa Aravindh\textsuperscript{\emph{4}}, Marko Huttula\textsuperscript{\emph{1}}, KeYuan Ma\textsuperscript{\emph{5}}, Fabian O. von Rohr\textsuperscript{\emph{6}}, Jean-Christophe Le Breton\textsuperscript{\emph{3}}, Thomas Jaouen\textsuperscript{\emph{3}}, and Claude Monney \textsuperscript{\emph{2}}\\[0.2cm]
\small \textsuperscript{\emph{1}} \textit{Nano and molecular systems research unit, University of Oulu, FI-90014 Oulu, Finland}\\
\small \textsuperscript{\emph{2}} \textit{Department of Physics and Fribourg Center for Nanomaterials, Universit$\mathrm{\acute{e}}$ de Fribourg; Fribourg CH-1700, Switzerland} \\
\small \textsuperscript{\emph{3}} \textit{Univ Rennes, CNRS, (IPR Institut de Physique de Rennes) - UMR 6251, F-35000 Rennes, France} \\
\small \textsuperscript{\emph{4}} \textit{Research Unit of Sustainable Chemistry, University of Oulu, FI-90014 Oulu, Finland} \\
\small \textsuperscript{\emph{5}} \textit{Max Planck Institute for Chemical Physics of Solids, 01187 Dresden, Germany} \\
\small \textsuperscript{\emph{6}} \textit{Department of Quantum Matter Physics, University of Geneva, CH-1211 Geneva, Switzerland} \\[0.4cm]

Corresponding authors: Kati.Asikainen@oulu.fi, claude.monney@unifr.ch
\end{center}
\hspace{0.3cm}

\begin{abstract}
Metal-assisted exfoliation of two-dimensional (2D) materials has emerged as an efficient route to isolating large-area monolayer crystals, yet the influence of the supporting metal substrate on their intrinsic properties remains poorly understood. Here, we demonstrate successful gold-assisted exfoliation of monolayer \ch{IrTe2} up to the millimeter scale. Angle-resolved photoemission spectroscopy (ARPES), combined with first-principles calculations, reveals that the low-energy electronic structure closely resembles that of a freestanding monolayer 1T-\ch{IrTe2}. We find that quasi-covalent hybridization together with substrate-induced strain leads to only modest modifications of the electronic bands. Although strain contributes to phase stability, it is essentially hybridization that drives the stabilization of the 1T-phase of the monolayer \ch{IrTe2} by suppressing stripe-ordered phase transitions. These results establish gold-assisted exfoliation as a robust route to prepare a large-area monolayer \ch{IrTe2} and highlight the role of metal-substrate interaction in engineering 2D materials with tailored structural phases. 

\end{abstract}

\section{Introduction}


Transition metal dichalcogenides (TMDs) are layered materials, featuring strong in-plane bonding with individual layers interacting through weak van der Waals (vdW) forces \cite{TMDs}. They have been extensively studied over the past 50 years, both experimentally and theoretically, because of the diversity of their electronic, optical, mechanical, and chemical properties. They exhibit rich phase diagrams, often involving charge density wave (CDW) phases \cite{CDW-0, CDW-1, CDW-2} coexisting or competing with other electronic phases, such as superconductivity or Mott insulating phases. Understanding and controlling phase transitions in TMDs could provide pathways for designing advanced electronic and quantum materials for various applications, such as phase-change memory and resistive switching devices \cite{appl2, appl3}, where switching between different phases enables new functionalities. 

Following the discovery of graphene by mechanical tape exfoliation in 2004 \cite{graphene}, it was soon realized that the thickness of TMD materials at the atomic layer level opens a new axis in their phase diagram. In practice, tape exfoliation places limitations on the investigation and application of atomically thin samples because exfoliated flakes typically have lateral sizes of tens of micrometers at most. More recently, new mechanical exfoliation methods based on metal substrates were proposed to overcome this limitation. Among them, gold-assisted exfoliation has been shown to produce mm-sized, even cm-sized, monolayer flakes \cite{Huang,Desai2016,velicky2018,Heyl2023,Panasci2024,WSe2,Pushkarna2023}. This technique relies on stronger hybridization between the gold substrate and the monolayer flake, which surpasses the vdW interlayer bonds of the TMDs and thereby favors monolayer exfoliation. The adhesion at the interface prevents the monolayer from breaking, enabling the production of large-area flakes. Additionally, substrate-induced strain, often due to the lattice mismatch with gold, is proposed to weaken interlayer coupling and facilitate monolayer exfoliation \cite{Heyl2023,Panasci2024,Ziewer2025,Velicky2020}. However, these effects can also be disruptive, as they may significantly alter the intrinsic physical properties of monolayer TMDs \cite{velicky2018,Bruix2016,Shao2019,Velicky,Pushkarna2023}.

Among TMDs, bulk IrTe$_2$ is known for its complex series of first-order structural phase transitions occurring below room temperature (RT). Already at 280 K, the crystal structure changes from a trigonal 1T phase ($P$-3$m$1) to a monoclinic one ($P$-1) involving a (5x1x5) superstructure \cite{Yang2012}, followed by a second transition at 180 K into an (8x1x8) phase \cite{Ko2015}. Low-temperature (LT) measurements revealed a (6x1x6) ground state \cite{Hsu-STM, Takubo2018}, which can also be stabilized via Se doping \cite{Oh2013} or uniaxial strain \cite{Nicholson2021}. Numerous ARPES studies have further elucidated the evolving electronic structure of IrTe$_2$ \cite{Ootsuki2013, Ootsuki2017, Qian2014, Ko2015, Blake2015, Lee2017a, Rumo, Nicholson2021, Bao2021, Rumo2021, Mizokawa2022}. Although the LT phases show reduced resistivity, bulk IrTe$_2$ remains metallic at all temperatures. Each of the phases below 280 K exhibits a substantial shortening of atomic bonds that produces Ir dimers with bonding and anti-bonding Ir 5$d$ states \cite{Pascut2014a, Pascut2014b}, as part of a more complex multicentre bond \cite{Saleh2020,Ritschel2022,Nicholson2024}. The number of dimers per unit cell determines the one-dimensional stripe pattern of LT phases with specific periodicity. Similarly, the reported results demonstrate various stripe-like phase transitions in 2D monolayer \ch{IrTe2} on different substrates. In a recent study, Hwang and co-workers produced monolayer IrTe$_2$ on bilayer graphene terminated 6$H$-SiC(0001) substrate \cite{Hwang} and showed that monolayer IrTe$_2$ is a semiconductor from RT down to 14 K due to a (2x1) structural reconstruction involving a fully dimerized lattice. Park \textit{et al.} \cite{Park2021} observed the coexistence of stripe order and superconductivity in monolayer IrTe2 on Si/\ch{SiO2} substrate, suggesting that stripes may enhance superconductivity. In contrast, Song \textit{et al.} \cite{IrTe2-Au-Raman} reports a competing relationship between CDW and superconductivity on the gold substrate. This study suggests that the high-temperature trigonal $P$-3$m$1 structure could be preserved locally in the superconducting regions. Furthermore, monolayer \ch{IrTe2} has been shown to exhibit thickness-dependent emergence of stripe-like charge orders on \ch{Al2O3} substrate, with the monolayer and bilayer notably showing no phase transitions \cite{IrTe2-Al2O3}. These studies indicate clear substrate-dependent phase behavior in monolayer \ch{IrTe2}, underscoring the critical role of interfacial interaction in dictating its structure and electronic phases. In particular, the behavior on metallic substrates has been poorly explored, motivating further investigations. 

In the present work, we produce a large mm-sized monolayer IrTe$_2$ on a polycrystalline gold substrate and characterize its atomic and electronic structure. Using scanning tunneling microscopy (STM) and ARPES, we show that there are no structural phase transitions to stripe-ordered phases below RT. Surprisingly, the low-energy electronic structure of our monolayer sample turns out to be very similar to that of a freestanding monolayer. Using a realistic atomic model of monolayer IrTe$_2$ with 1T phase on Au(111) within density functional theory (DFT), we show that hybridization and strain are the primary factors modifying the electronic structure of the monolayer IrTe$_2$, introducing only minor changes. Phonon dynamics calculations demonstrate that the interaction with gold is crucial for stabilizing the 1T atomic structure. Together with previous works \cite{Hwang,Park2021,IrTe2-Au-Raman,IrTe2-Al2O3}, this work exemplifies how much the dielectric and electronic environments of a substrate impact the physical properties of atomically thin samples.

\section{Results}

\subsection{Structural characterization of Au-supported \ch{IrTe2} monolayer}
\begin{figure}[h!]\centering
\includegraphics[width=1\linewidth]{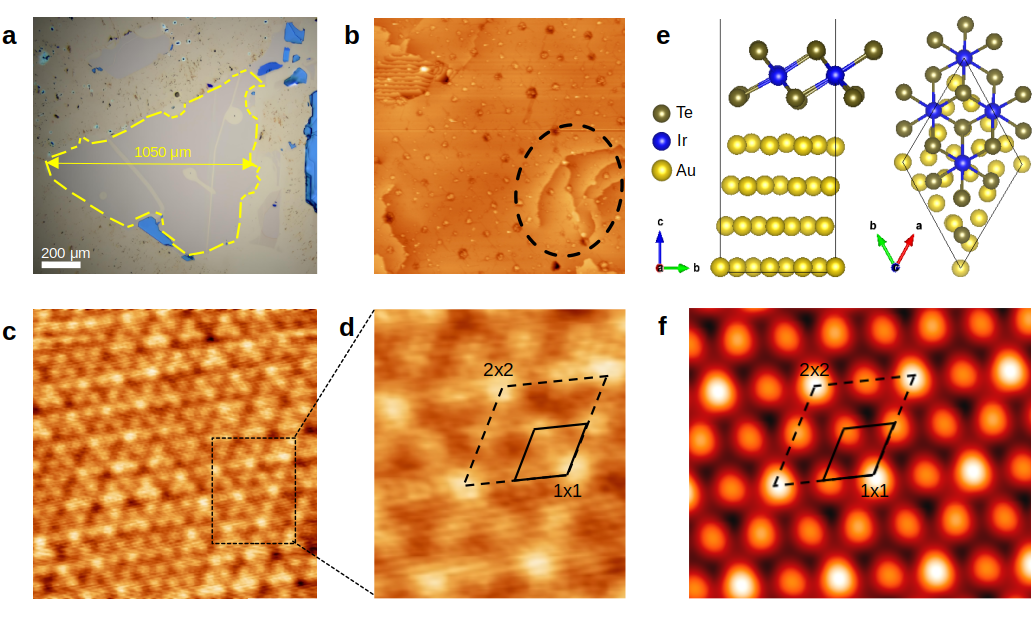}
\caption{\textbf{Crystal structure and characterization of \ch{IrTe2} monolayers:} \textbf{a} Optical image of \ch{IrTe2} monolayer flake (outlined by dashed yellow line) on a polycrystalline Au substrate. \textbf{b} Large-scale STM topograpy image ($200 \times 200\ \mathrm{nm}{}^2$, I=0.2 nA, V=$-0.3$ V, plane correction, flattened), with step edges and terraces highlighted with a dashed black circle. \textbf{c} STM topography image with atomic resolution ($6.172 \times 6.172\  \mathrm{nm}{}^2$, I=1.3 nA, V=$-0.005$ V, plane correction, flattened, Gaussian smoothed). The region marked by a dotted rectangle is enlarged in \textbf{d}. Black solid and dashed rhombuses indicate the (1 $\times$ 1) unit cell of monolayer \ch{IrTe2} and (2 $\times$ 2) modulation, respectively. \textbf{e} Side and top views of the optimized \ch{IrTe2}-Au structure. \textbf{f} DFT-simulated STM image of \ch{IrTe2}-Au, with $-0.3$ bias voltage, extracted 4 \AA \ above the monolayer surface, predicting the observed (2 $\times$ 2) modulation, consistent with \textbf{d}.}\label{Fig1}
\end{figure}

\ch{IrTe2} samples exfoliated on polycrystalline gold were first characterized by optical microscopy. Monolayer \ch{IrTe2} flakes appear as transparent-like regions (Fig. \ref{Fig1}a) with slightly darker contrast relative to the bare Au substrate. The observed monolayers typically range from 100 to $\SI{600}{\micro\meter}$ in size, some flakes reaching up to $\SI{1000}{\micro\meter}$, demonstrating the feasibility of producing millimeter-scale large-area \ch{IrTe2} monolayers. 

Figure \ref{Fig1}b shows a large-scale STM image of our \ch{IrTe2} monolayer sample taken at 4.5 K. The smooth surface exhibits occasional terraces separated by step edges.
The measured step height is approximately 0.24 nm (Supplementary note 1 and Fig. S1), consistent with the intrinsic height of a single Au(111) atomic plane. This confirms the successful exfoliation of the monolayer \ch{IrTe2}, as no additional step heights associated with the underlying layers were observed. It also suggests a dominant presence of Au grains with (111) orientation within the polycrystalline substrate, which is expected from thermodynamics. The flakes exhibit excellent uniformity and surface morphology as revealed by scanning electron microscopy (SEM) imaging (Supplementary note 2 and Fig. S2). Furthermore, atomic force microscopy (AFM) height profile measurements of \ch{IrTe2} line patterns show a monolayer thickness of approximately 0.65 nm, typical of TMD monolayers, further providing confidence in our findings (Supplementary note 2 and Fig. S2).

The fine atomic structure was identified from the atomically-resolved STM image of the monolayer \ch{IrTe2} (Fig. \ref{Fig1}c), showing the characteristic hexagonal symmetry of the 1T phase, with minor deviations from the ideal symmetry. Notably, we observe a weak local (2 $\times$ 2) modulation of the electronic density visible through slight contrast differences (Fig. \ref{Fig1}d). To analyze whether this indicates a new structural phase, we computationally modeled a (2 $\times$ 2) monolayer \ch{IrTe2} on the Au(111) surface, hereafter referred to as \ch{IrTe2}-Au. The choice of Au(111) surface is based on the assumption that (111)-oriented grains are prevalent in the polycrystalline gold substrate, supported by STM. Figure \ref{Fig1}e shows the optimized structure of the supercell model. The DFT-simulated STM image reflects the (2 $\times$ 2) modulation seen in the experiments, with one atom appearing brighter and three dimmer per unit cell (Fig. \ref{Fig1}f). Since the monolayer maintains its 1T hexagonal symmetry in DFT calculations, the contrast variations can be attributed to a mild structural distortion at the monolayer-substrate interface, particularly involving pronounced vertical displacements that modify the surface charge density (Supplementary note 3 and Fig. S3). Consequently, the (2 $\times$ 2) modulation does not indicate a true phase transition but rather an interfacial effect. Therefore, we rule out the occurrence of any low-temperature phases with stripes of Ir dimers and any structural reconstruction of the 1T phase. This is in stark contrast to measurements performed on bulk \ch{IrTe2} \cite{Hsu-STM}, as well as on monolayer \ch{IrTe2} on bilayer graphene \cite{Hwang}.

\subsection{Low-energy electronic structure of Au-supported \ch{IrTe2} monolayer}
\begin{center}
\begin{figure}[h!]\centering
\includegraphics[width=0.95\linewidth]{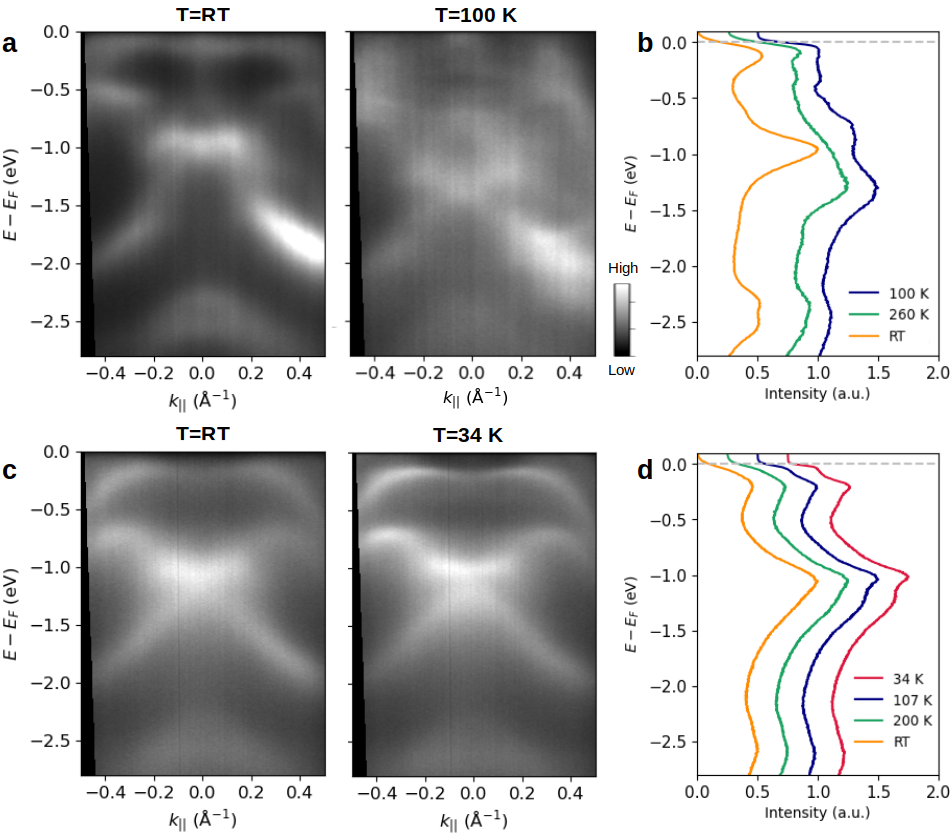}
\caption{\textbf{Electronic structure of bulk and monolayer \ch{IrTe2}:} (\textbf{a}, \textbf{c}) ARPES intensity maps of bulk \ch{IrTe2} along L-A-L high symmetry direction and monolayer \ch{IrTe2} along M-$\Gamma$-M, respectively. Panel \textbf{b} shows the corresponding EDCs of three phases of bulk \ch{IrTe2} while panel \textbf{d} displays EDCs of monolayer \ch{IrTe2} at different temperatures. The Fermi level is indicated with a dashed line in panels \textbf{b} and \textbf{d}. The EDCs were obtained by integrating over $\pm 0.07$\ \AA$^{-1}$ around A in the bulk and $\Gamma$ in the monolayer.}\label{Fig2}
\end{figure}
\end{center}

Let us now focus on the reasons why no structural phase transitions occur in our samples. For this purpose, we performed ARPES measurements of their low-energy electronic structure. For comparison, Fig. \ref{Fig2}a shows the electronic structure of bulk \ch{IrTe2} at RT and in the lowest temperature phase driven by Ir dimerization. While the distinction between the first and second low-temperature phases is subtle in the ARPES spectra, the corresponding energy distribution curves (EDCs) in Fig. \ref{Fig2}b allow for a clear identification of the distinct spectral features associated with all three phases. Our measurements nicely reproduce the results reported by Rumo \textit{et al.} \cite{Rumo}, providing a benchmark for the analysis of the monolayer. 

\clearpage

Figure \ref{Fig2}c displays the electronic structure of the Au-supported \ch{IrTe2} monolayer at RT. We observe well-defined electronic bands, distinct from the bulk spectrum. In particular, the absence of a surface resonance state around -0.5 eV at $\Gamma$ strongly indicates the monolayer character, since this state is characteristic of the bulk \ch{IrTe2} \cite{Rumo}. Upon cooling from RT down to 34 K, we find that the ARPES band dispersions remain essentially unchanged (Fig. \ref{Fig2}c). The fine structure becomes more pronounced at lower temperatures (Fig. \ref{Fig2}c), but these small spectral changes can be attributed to a reduced thermal broadening. Together with STM, these results further confirm the absence of stripe-ordered phase transitions in the Au-supported \ch{IrTe2} monolayer down to 4.5 K.

\subsection{DFT insight into electronic structure and substrate interaction}

\begin{figure}[h!]\centering
\includegraphics[width=0.86\linewidth]{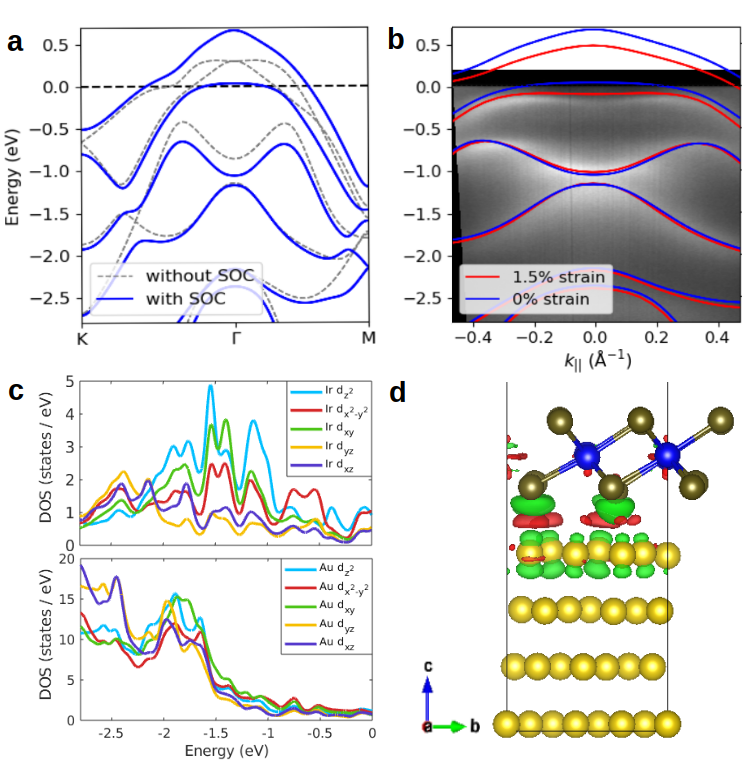}
\caption{\textbf{DFT-calculated electronic properties of monolayer \ch{IrTe2}:} \textbf{a} DFT band structure of freestanding monolayer \ch{IrTe2} with and without SOC. \textbf{b} ARPES intensity map at 34 K overlaid with DFT band structure for the undistorted case (from \textbf{a}) and 1.5\% biaxially strained monolayer \ch{IrTe2}, both calculated including SOC. \textbf{c} Density of states (DOS) of Ir and Au \textit{d} orbitals in the \ch{IrTe2}\text{-}Au model, calculated with SOC. \textbf{d} Charge density difference plot for the \ch{IrTe2}-Au model, with an isosurface value of 0.003 e/\AA³. Red refers to electron accumulation, and green to electron depletion.}\label{Fig3}
\end{figure}

We carried out DFT calculations to provide theoretical support for the ARPES measurements and to gain deeper insights into the absence of phase transitions. We began by calculating the DFT band structure of an undistorted freestanding monolayer \ch{IrTe2} with and without spin-orbit coupling (SOC) (Fig. \ref{Fig3}a). SOC lifts band degeneracies at certain momenta, resulting in gap openings. Inclusion of SOC leads to excellent agreement with ARPES data (Fig. \ref{Fig3}b), underscoring its importance in accurately capturing the electronic structure. The most notable deviation concerns the position of the bands near the Fermi level. DFT predicts two hole-like bands with maxima at $\Gamma$ crossing the Fermi level, while in the ARPES spectrum, the band with its maximum closer to the Fermi level is flat and pushed below it. These discrepancies point to substrate-induced effects in tuning the electronic structure, prompting further investigations. Motivated by STM observations suggesting a mild lattice distortion, we modeled a strain in the freestanding monolayer (Supplementary note 4 and Fig. S4) and found that it modifies the bands toward the experimentally observed dispersion. Figure \ref{Fig3}b also shows the band structure (with SOC) of 1.5\% biaxially strained monolayer \ch{IrTe2} overlaid on the ARPES spectrum. The primary effect is to shift the two hole-like bands downward in energy, pushing the lower one fully below the Fermi level, improving the agreement between DFT and ARPES. The band structure of the strained monolayer also shows enhanced agreement with the STS data (Fig. S5), supporting the strain-driven changes in the electronic structure.  

The simple model of a freestanding monolayer \ch{IrTe2} with strain reproduces remarkably well the electronic structure measured by ARPES down to 1.5 eV below the Fermi level. This is somewhat surprising, given that there must be a substantial interfacial coupling with the Au substrate, necessary for efficient gold-assisted exfoliation. To gain more understanding of the interfacial interaction, we calculated the electronic structure of \ch{IrTe2}-Au (Figure \ref{Fig1}e). 
Unlike STM, which can probe nanometer-size domains (Fig. \ref{Fig1}c), our ARPES measurements average over an area of about $500 \times 500~\mathrm{\upmu m^2}$, involving predominantly Au(111) grains with different in-plane orientations. Therefore, since our model represents only a local region observed in the STM, we focus on analyzing the projected density of states (DOS) only, shown in Fig. \ref{Fig3}c. The projected bulk states of Au increase drastically at $-1.5$ eV. We found a pronounced overlap between the out-of-plane Ir and Au \textit{d} orbitals, dominated by $d_{yz}$ and $d_{xz}$ orbitals below this binding energy, where a stronger monolayer-substrate interaction is expected \cite{Takeuchi1990, Miwa2015, Bruix2016}. This is indicative of quasi-covalent hybridization at the interface and is somewhat visible in our ARPES data, where the bands appear blurred at higher binding energies (Fig. \ref{Fig2}c). In contrast, lower-binding-energy states lie within the projected band gap of Au along the (111) direction \cite{Takeuchi1990,Miwa2015}. Although we observed slight intensity losses, namely hybridization gaps around $-0.9$ eV at $\pm$ 0.2 \AA$^{-1}$ (Fig. \ref{Fig2}c), it is generally expected that the interaction between Au and \ch{IrTe2} is weak in this energy region due to reduced orbital overlap \cite{Miwa2015}. 

Let us now quantify the strength of the hybridization. As shown in Fig. \ref{Fig3}\textbf{d}, charge redistribution is concentrated at the interface, which is expected when a monolayer is placed on a metallic surface. The adhesive energy in the \ch{IrTe2}-Au model was calculated to be 0.074 eV/\AA², which is comparable to the value reported by Huang \textit{et al.} for \ch{IrTe2} (0.0680 eV/\AA²) \cite{Huang}. In contrast, the adhesive energy of \ch{MoS2} on the Au(111) surface is 0.0397 eV/\AA² \cite{Huang}, roughly half that of \ch{IrTe2}. Bader analysis indicates that electrons are transferred to the Au surface, resulting in hole doping of \ch{IrTe2}, consistent with the work function alignment of monolayer \ch{IrTe2} and Au(111) surface ($\Phi_{\mathrm{IrTe_2}}=4.88$ eV and $\Phi_{\mathrm{Au}}=5.21$ eV). The calculated net charge transfer of 7.6 $\times 10^{13}$ \textit{e}/cm${}^2$ across the interface further shows a stronger monolayer-substrate interaction compared to \ch{MoS2} on Au, where the transfer is reported to be below 2.0 $\times 10^{13}$ \textit{e}/cm${}^2$ \cite{BarZiv2019, Kim2021}. These findings reflect a significant interaction between the monolayer \ch{IrTe2} and the Au substrate, which is central to enabling exfoliation, and underscore the influence of interfacial coupling on the monolayer's properties.

\section{Discussion}

\begin{figure}[h!]\centering
\includegraphics[width=1\linewidth]{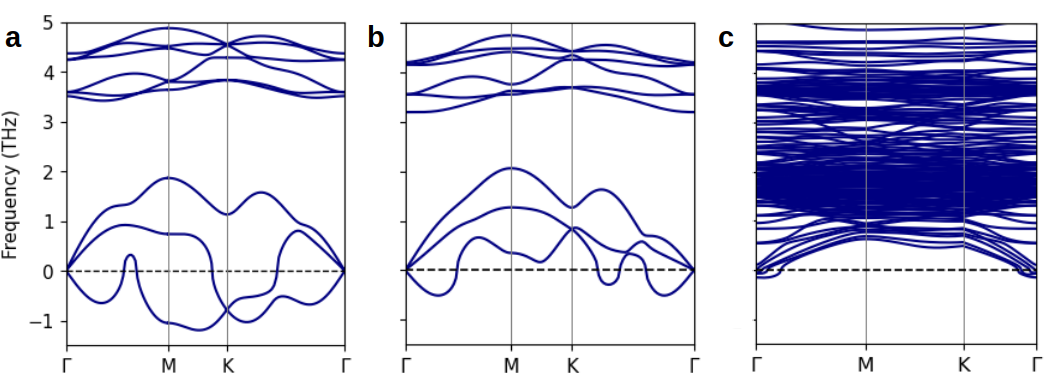}
\caption{\textbf{Phonon calculations:} Phonon band structure of a) unstrained and b) 1.5\% biaxially strained monolayer \ch{IrTe2}, and c) the \ch{IrTe2}-Au model, computed with SOC.}\label{Fig4}
\end{figure}


Although the band structure of the Au-supported \ch{IrTe2} monolayer appears remarkably similar to that of the freestanding one, strain and quasi-covalent hybridization are found to alter its electronic properties. However, it is not yet clear whether these factors prevent phase transitions in \ch{IrTe2} monolayers. To address this, we performed phonon calculations using density functional perturbation theory (DFPT) to examine their influence on the stability of monolayer \ch{IrTe2}. For this, we considered freestanding and 1.5\% biaxially strained monolayers and an Au-supported one (\ch{IrTe2}\text{-}Au). The thermodynamic instability of a freestanding \ch{IrTe2} monolayer is well established from phonon band structure calculations \cite{Hwang, Na-IrTe2, Song}. In line with previous reports, our phonon calculations reveal the presence of imaginary frequencies (Fig. \ref{Fig4}a). While the applied strain tends to harden the phonons, it is not sufficient to fully stabilize the monolayer (Fig. \ref{Fig4}b). This contrasts with the findings of Song \textit{et al.} \cite{Song}, who reported that a small biaxial strain could completely eliminate structural instability. While this discrepancy likely arises from different computational methods, our results show the same qualitative trend, which is that strain mitigates structural instability. In contrast, when the monolayer is placed on the Au surface, the imaginary frequencies are suppressed. This shows that the monolayer \ch{IrTe2} in the 1T phase is thermodynamically stable on a gold substrate, consistent with experimental observations. Although small imaginary frequencies remain near $\Gamma$ in the acoustic modes, they were identified to originate from numerical noise rather than physical instability (Supplementary note 5 and Fig. S6). Thus, the findings indicate that the interaction with gold, particularly hybridization, along with strain, plays an active role in stabilizing the 1T structure of the \ch{IrTe2} monolayer against any stripe-ordered phase transitions.

In conclusion, by combining STM, ARPES, and first-principles DFT calculations, we have demonstrated that, unlike the ($2 \times 1$) dimerized structure observed on an insulating substrate \cite{Hwang}, Au-supported \ch{IrTe2} monolayer retains the typical hexagonal 1T bulk crystal structure from room temperature down to 4.5 K. We have shown that hybridization is the key factor in stabilizing the structure, thus inhibiting any stripe-ordered structural phase transitions. Although the electronic band structure is modified by these interactions, the overall low-energy dispersions closely resemble to those of the freestanding \ch{IrTe2} monolayer. Our findings provide important insight into the role of the underlying substrate in determining phase stability, and offers both understanding and guidance for the design of future 2D material-based devices. 

\section{Methods}

\subsection{Sample preparation}
The samples were prepared on \ch{SiO2} (90 nm)/p++Si substrates. A 2 nm chromium adhesion layer followed by a 10 nm polycrystalline gold layer was sputter deposited onto the substrate. The \ch{IrTe2} surface was refreshed with Scotch tape to ensure a clean interface and immediately brought in contact with the freshly deposited gold layer using a thermal release tape (TRT). It was then released from the TRT by heating the substrate to 200 $^{\circ}$C. The resulting \ch{IrTe2}/Au/Cr/\ch{SiO2}/Si sample was introduced into ultra-high vacuum ($\sim 10^{-8}$ mbar) at room temperature, where Scotch tape was used to peel off the bulk crystal, leaving behind large Au-supported \ch{IrTe2} monolayers. 

\subsection{Scanning probe microscopy measurements}

STM measurements were performed at a base pressure better than $5\times 10^{-11}$ mbar, and constant current STM images were recorded at 4.5 K using an Omicron low-temperature (LT)-STM, with a bias voltage applied to the sample. The differential conductance dI/dV curves (STS) were recorded with an open feedback loop using the standard lock-in method.

We performed AFM measurements using an NT-MDT NTEGRA microscope operating in non-contact tapping mode, and SEM imaging with a Hitachi FlexSEM 1000 II microscope operated at 3 kV in low-vacuum mode without conductive coating to characterize the surface morphology, homogeneity, and thickness of \ch{IrTe2} flakes. All measurements were carried out at RT.

\subsection{ARPES measurements}
Temperature-dependent ARPES investigations were carried out using a Scienta DA30 photoelectron analyzer (R8000) with a base pressure better than $3\times 10^{-11}$ mbar. Monochromatized He $I_\alpha$ radiation with $h\nu = 21.22$ eV with a final spot size diameter of 750 $\mu$m was used. The total energy resolution was about 10 meV and cooling of the sample was carried out at rates $<$ 5 K/min to avoid thermal stress. Each measurement was preceded by a break of at least 15 min, to ensure thermalization. Accordingly, the error on the absolute sample temperature was estimated to be below 5 K.

\subsection{DFT calculations}
First-principles DFT calculations were performed using the Vienna ab initio simulation package (VASP) \cite{vasp1, vasp2, vasp3}. Our calculations employed the generalized gradient approximation (GGA) \cite{GGA} for exchange-correlation functional, together with the projector augmented-wave (PAW) method \cite{PAW}. The energy cutoff for the plane wave basis set was set to 650 eV and Gaussian smearing of 0.05 eV was used for Brillouin-zone integrations. 
We modeled the freestanding monolayer \ch{IrTe2} using the hexagonal 1T unit cell and gold surface using four-layer Au(111) slab. The heterostructure was created by combining the (2 $\times$ 2) monolayer \ch{IrTe2} cell with Au(111) slab, labeled as \ch{IrTe2}-Au (Supplementary note 6). A vacuum thickness of around 15 \AA \ was added to avoid spurious interaction between periodic images. All structures were fully relaxed until the forces were less than 0.001 eV/\AA, while the bottom layer of the Au(111) was kept fixed in the heterostructure. We use a k-mesh of ($9 \times 9 \times 1$) for monolayer \ch{IrTe2} and ($4 \times 4 \times 1$) for \ch{IrTe2}-Au according to Monkhorst-Pack scheme \cite{MP}. In \ch{IrTe2}-Au, the dispersion correction within the Tkatchenko-Scheffler scheme (DFT-TS)\cite{TS} was applied to describe the van der Waals interaction at the interface (Supplementary note 7 and Table S1). Electronic structure and phonon calculations were performed including SOC. To analyze the charge density distribution, we calculated the charge density difference according to $\Delta \rho = \rho(\mathrm{IrTe_2\text{-}Au}) - \rho(\mathrm{IrTe_2}) - \rho(\mathrm{Au})$, where $\rho(\mathrm{IrTe_2\text{-}Au})$, $\rho(\mathrm{IrTe_2})$ and $\rho(\mathrm{Au})$ are the charge densities of the heterostructure, monolayer \ch{IrTe2}, and Au(111) surface, respectively. Bader analysis was performed using the Henkelman code \cite{Bader1, Bader2} to quantitatively evaluate charge transfer. Phonon calculations were performed within the DFPT \cite{dfpt}, including SOC. Harmonic force constants and phonon band structures were obtained using the Phonopy \cite{phonopy} code. For \ch{IrTe2}\text{-}Au, the k-mesh was reduced to $3 \times 3 \times 1$ to balance computational efficiency in phonon calculations.

\section*{Acknowledgements}

This work was supported by the European Union’s Horizon Europe research and innovation program under the Marie Sk\l odowska-Curie grant agreement no. 101081280, and by the Swiss National Science Foundation (SNSF) Grant No.10.000.782. T. J. acknowledges the support of the French National Research Agency (ANR) (MOSAICS project, ANR-22-CE30-0008). Computational resources were provided by the CSC - IT Center for Science, Finland.

\section*{Declaration of competing interest}
The authors declare no competing financial interests.

\end{document}


\begin{center}
{\Large\textbf{Supplementary information}} \\[8pt]
{\large\textbf{Suppression of stripe-ordered structural phases in monolayer \ch{IrTe2} by a gold substrate}} \\[8pt]
Kati Asikainen, Fr$\mathrm{\acute{e}}$d$\mathrm{\acute{e}}$ric Chassot, Baptiste Hildebrand, Aymen Mahmoudi, Joel Morf, Mahault Berset, Pascal Turban, Matti Alatalo, S. Assa Aravindh, Marko Huttula, KeYuan Ma, Fabian O. von Rohr, Jean-Christophe Le Breton, Thomas Jaouen, and Claude Monney\\[0.3cm]

Corresponding authors: Kati.Asikainen@oulu.fi, claude.monney@unifr.ch
\end{center}

\clearpage

\noindent\textbf{Supplementary note 1: Determining step heights using STM}\\

\begin{figure}[h!]\centering
\includegraphics[width=0.92\linewidth]{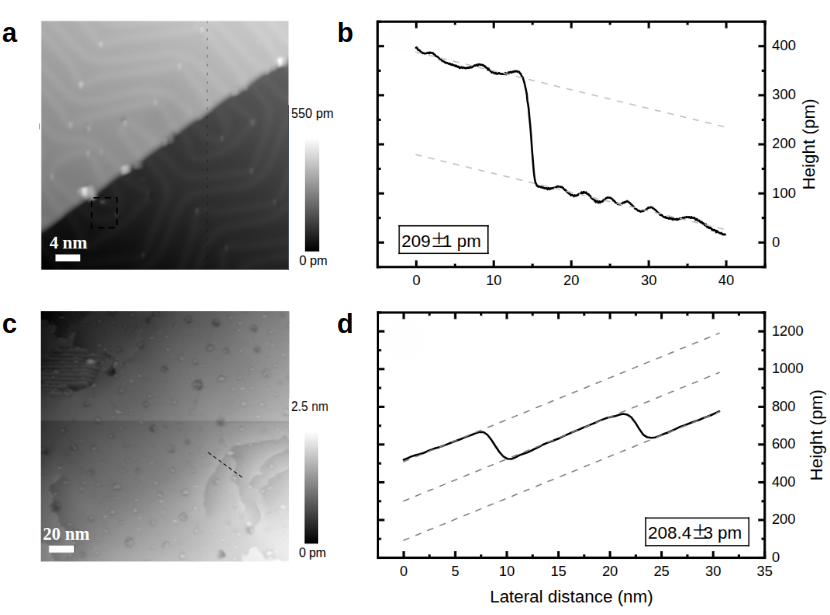}
\caption{\textbf{a} STM topographic image (40 $\times$ 40 $\mathrm{nm}{}^2$, I = 1 nA, V = $-1$ V) showing a terrace on a gold-only region. \textbf{b} Height profile along the dashed black line in \textbf{a}, indicating a step height of 209 pm. \textbf{c} Large-scale STM topographic image (200 $\times$ 200 $\mathrm{nm}{}^2$, I $= 0.2$ nA, V $=-0.3$ V) of monolayer \ch{IrTe2} on a gold substrate. \textbf{d} Height profile along the dashed black line in \textbf{c}, revealing a double step on monolayer \ch{IrTe2} with a height of 208.4 pm. The step heights were determined by fitting the terraces using first-order polynomial leveling, with the fitted levels shown as gray dashed lines.}\label{FigS1}
\end{figure}

We characterized the step edges of Au-supported \ch{IrTe2} monolyares using Au(111) terraces as a calibration reference. Fig. S\ref{FigS1}a, c shows raw (unflattened) STM scans of Au(111) and monolayer \ch{IrTe2} flake regions with visible terraces. The height profiles in Fig. S1b, d are consistent, confirming that monolayer \ch{IrTe2} follows the same step structure as the underlying gold substrate. The apparent step height obtained from our STM on Au(111) terraces was approximately 0.21 nm, slightly less than the well-established value of 0.24 nm \cite{Barth}. This systematic underestimation arises from instrument-specific calibration and measurement conditions, which commonly affect STM height accuracy. Therefore, a scaling correction factor of $\sim$1.2 was applied to all height data to ensure consistency with the reference value. With the correction factor applied, the step height measured in the monolayer \ch{IrTe2}, shown in \textbf{d}, aligns with the intrinsic Au(111) terrace height, as reported in the main text. The shown profiles in Fig. S\ref{FigS1} represent examples from a comprehensive statistical analysis of step height measurements performed across multiple terraces on different zones and various step configurations, validating the uniformity of the step morphology.

\clearpage

\noindent\textbf{Supplementary note 2: AFM and SEM characterization of Au-supported \ch{IrTe2} monolayer}\\

\begin{figure}[h!]\centering
\includegraphics[width=0.81\linewidth]{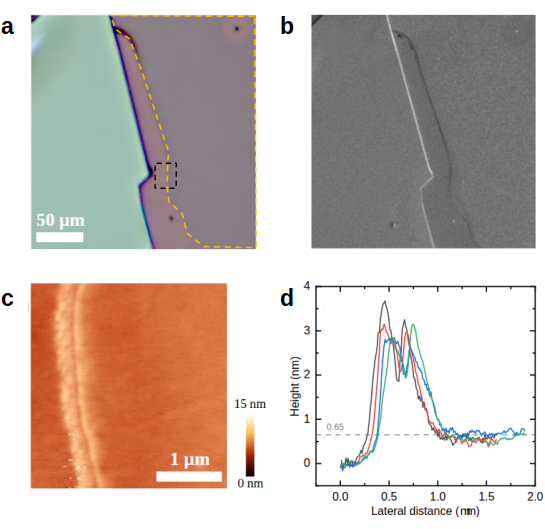}
\caption{\textbf{a} Optical image of monolayer \ch{IrTe2}, outlined by the yellow dashed line, and \textbf{b} the corresponding SEM image of the same region. \textbf{c} AFM topography of the substrate-monolayer interface region marked by the dashed square in \textbf{a}. AFM height profiles taken across the interface, confirming a monolayer thickness of approximately 0.65 nm.}\label{FigS-AFM}
\end{figure}

AFM and SEM scans in Fig. \ref{FigS-AFM} reveal a flat and homogeneous surface, consistent with the formation of a monolayer \ch{IrTe2}. The measured average step height is about 0.65 nm $\pm$ 0.1 nm, corresponding to a single \ch{IrTe2} layer. To the best of our knowledge, the exact monolayer thickness of \ch{IrTe2} has not yet been reported in the literature. However, this measured value is comparable to the typical monolayer thicknesses observed in transition-metal dichalcogenides \cite{Li2015,WS22013}. The non-uniform thickness observed can be ascribed to the roughness of the underlying gold substrate and the possible presence of a thin oxide layer. The spike observed at the step frontiers in the AFM topography is most likely associated with surface contamination or dust adsorption that occurred after the sample was exposed to ambient conditions.

\clearpage

\noindent\textbf{Supplementary note 3: Rumpling at the interface of monolayer \ch{IrTe2} and Au surface}\\

\begin{figure}[h!]\centering
\includegraphics[width=0.5\linewidth]{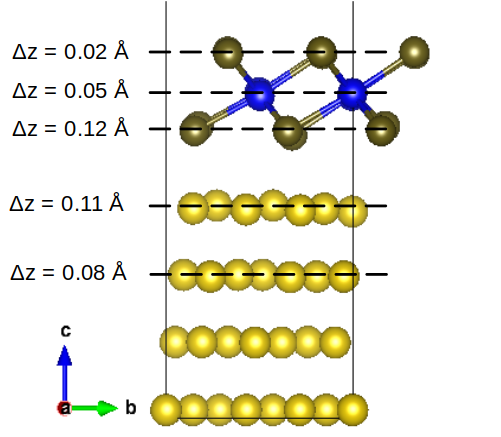}
\caption{Rumpling shown for the selected atomic planes in the \ch{IrTe2}\text{-}Au.}\label{Fig-rumpling}
\end{figure}

Due to the small lattice mismatch (See Supplementary note 6) and atomic alignments, mild atomic displacements are observed across the interface upon structural relaxation. We defined the rumpling of an atomic plane as $\Delta z = max(z)-min(z)$, where $z$ are the \textit{z}-coordinates of atoms within the same plane, which captures the vertical distortion (Fig. \ref{Fig-rumpling}). While the top Te atoms remain relatively flat, vertical atom displacements are more pronounced in the interfacial Te layer and the topmost Au layer. This suggests that the contrast variation in the STM with atomic resolution is driven by interfacial coupling, which modulates the surface electronic density.

\clearpage

\noindent\textbf{Supplementary note 4: Applying strain in monolayer \ch{IrTe2}}\\

\begin{figure}[h!]\centering
\includegraphics[width=0.95\linewidth]{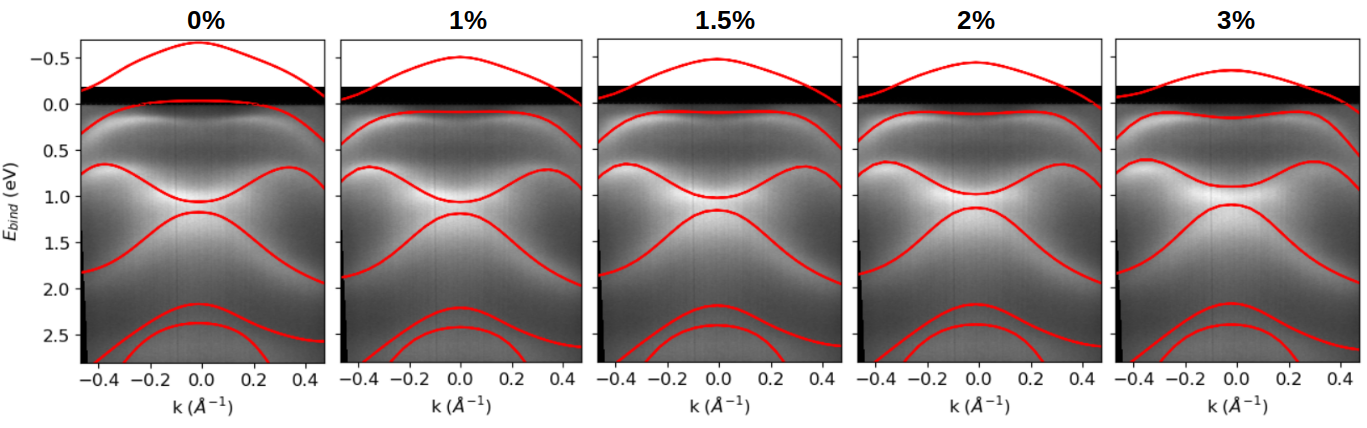}
\caption{ARPES intensity map of monolayer \ch{IrTe2} at 34 K overlaid with DFT band structures calculated under different biaxial strain values.}\label{Fig-strain}
\end{figure}

\begin{figure}[H]\centering
\includegraphics[width=0.93\linewidth]{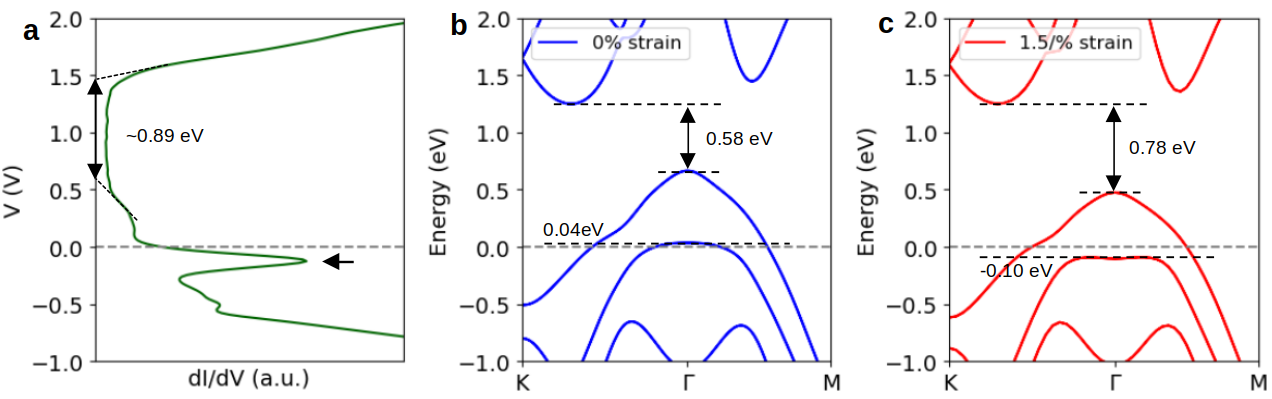}
\caption{\textbf{a} STS spectrum of monolayer \ch{IrTe2} from -1.0 V to 2.0 V, with the peak at -0.13 V indicated by a black arrow. (\textbf{b}, \textbf{c}) The DFT band structure of undistorted and 1.5\% biaxially strained monolayer \ch{IrTe2}, respectively. The maximum of the second highest band at $\Gamma$, corresponding to the peak in \textbf{a}, is marked by a black dashed line. The up-down arrow represents the band gap in both STS and DFT data.}\label{Fig-STS}
\end{figure}

Although STM in Fig. 1c suggests a mild structural deviation from the ideal hexagonal symmetry in monolayer \ch{IrTe2}, we were unable to unambiguously determine the exact nature of the distortion. Moreover, the underlying polycrystalline substrate is likely to influence the structure, leading to spatially non-uniform strain throughout the monolayer \ch{IrTe2}. Therefore, we considered a simple representative case, biaxial strain, in our calculations to demonstrate the strain effect and compare it with ARPES. Fig. \ref{Fig-strain} shows the electronic structure of monolayer \ch{IrTe2} under 0\%, 1\%, 1.5\%, 2\%, and 3\% biaxial strain. The comparison shows that moderate strain improves the agreement between DFT and ARPES, whereas higher strain values cause larger shift in the bands, reducing consistency with the experimental data. The 1.5\% biaxially strained case provides reasonable agreement with ARPES, and is also in closer agreement with STS data (Fig. \ref{Fig-STS}), thus serving an approximate representation of the strain effect supported by experimental evidence.


\hspace{1cm}


\noindent\textbf{Supplementary note 5: Analysis of small imaginary frequencies in the phonon band structure}\\

\begin{figure}[h!]\centering
\includegraphics[width=1\linewidth]{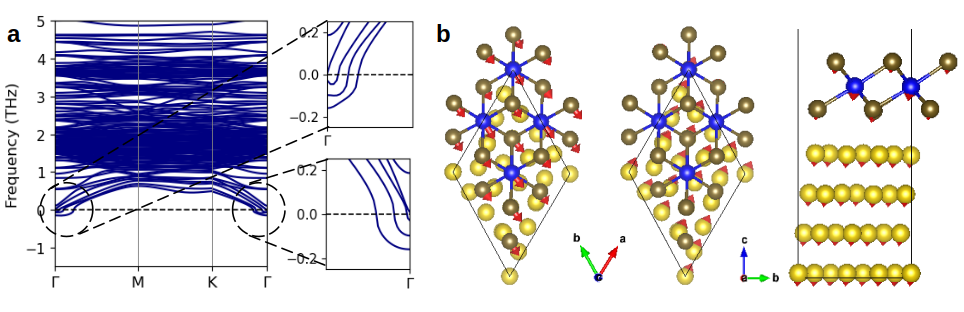}
\caption{a) Zoom-in on the three acoustic modes at $\Gamma$. b) Corresponding eigenvectors (red arrows) of each mode show uniform direction and magnitude, indicating rigid translations and supporting interpretation of small imaginary frequencies as numerical noise.}\label{Fig-phonon}
\end{figure}

Figure 4c shows the phonon band structure of \ch{IrTe2}\text{-}Au in the presence of SOC, showing some small imaginary frequencies at $\Gamma$. We analyzed the origin of these frequencies to determine whether they reflect real instability or arise from numerical artifacts \cite{Roy2024}. As a result, imaginary frequencies occur solely near the $\Gamma$ point in the three acoustic branches (Fig. \ref{Fig-phonon}a). The corresponding eigenvectors exhibit a rigid collective motion typical of acoustic modes (Fig. S\ref{Fig-phonon}b). This strongly supports the conclusion that the imaginary frequencies most likely arise from numerical inaccuracies rather than from a true dynamical instability. 

Because DFPT phonon calculations involve second derivatives of total energy, they are known to be sensitive to numerical parameters. Since using a denser k-grid in the calculation did not suppress artifacts at $\Gamma$, we attribute them to the finite supercell size. Vibrations involve long-range interactions, and therefore it is important to select an appropriate supercell size to achieve reliable convergence and accurate phonon modes. In a small supercell, vibrations can be artificially truncated by the periodic boundary condition, leading to numerical noise \cite{phonon1, phonon2}. Accordingly, within our computational setup, we focused on analyzing the origin of the imaginary frequencies, which we ultimately identified as numerical artifacts.\\

\vspace{1.5cm}



\noindent\textbf{Supplementary note 6: Constructing \ch{IrTe2}\text{-}Au supercell}\\

The in-plane lattice parameters of the optimized monolayer \ch{IrTe2} and Au(111) surface unit cells were $a=b=3.89$ \AA \ and $a=b=2.95$ \AA, respectively. To minimize lattice mismatch in the calculations, Au(111) unit cell was transformed using the following matrix, 
\begin{equation*}
\begin{bmatrix}
1 & 3 & 0  \\
3 & 2 & 0  \\
0 & 0 & 1  
\end{bmatrix} 
\end{equation*}
to match the lattice constant of (2$\times$2) monolayer \ch{IrTe2} cell. The resulting lattice parameters were 7.80 \AA \ and 7.77 \AA, respectively, giving an initial lattice mismatch of 0.38\% using the formula $\frac{a_{\mathrm{Au}}-a_{\mathrm{IrTe_2}}}{a_{\mathrm{Au}}}$. This minimizes artificial strain and allows for more realistic modeling of the combined system.    

\vspace{1.2cm}

\noindent\textbf{Supplementary note 7: Selection of the van der Waals method}\\

To account for dispersive forces and improve the description of interface interaction, we considered a vdW correction in our calculations for the \ch{IrTe2}\text{-}Au supercell. Because no single vdW method is universally applicable, we benchmarked several vdW correction schemes and functionals against experiments to identify the most appropriate one for our system. We considered the following dispersion methods: Tkatchenko-Scheffler method (DFT-TS) \cite{TS}, DFT-D3 \cite{D3}, vdW-DF \cite{DF}, and optB86b-vdW \cite{optB}. In addition, we optimized the structure with SOC to observe its impact on structural parameters. All of these were tested within the GGA framework. The structural parameters of interest are listed in Table S1. 

From the experiments, we measured a terrace height of approximately 2.4 \AA (Supplementary note 1 and Fig. S1), consistent with the lattice spacing of Au(111) surface. Standard GGA optimization (without vdW or SOC) yields an average lattice spacing of 2.46 \AA \ for the Au(111) surface in \ch{IrTe2}\text{-}Au, slightly overestimating the experimental value. We found that DFT-TS, DFT-D3 and optB86b-vdW methods produce lattice spacing in good agreement with the experiments ($\sim$2.4 \AA), highlighting the importance of including vdW interactions for structural modeling. Among the tested methods, we selected DFT-TS for our calculations. Although DFT-D3 and optB86b-vdW are generally considered among the most accurate vdW methods, DFT-TS predicts nearly identical structural geometry, while simultaneously being more suitable for our investigations due to its lower computational cost and compatibility with density functional perturbation theory (DFPT), enabling phonon calculations within the same framework. 
While SOC plays a crucial role in accurately predicting the electronic structure, it has limited influence on total energy minimization, and thus structural parameters, offering only little enhancement over GGA. There are also slight variations in the monolayer height and interlayer distance across the different functionals, yet they are largely comparable. The original vdW-DF clearly overestimates the experimental lattice spacing, in line with the observations of Klime$\mathrm{\check{s}}$ \textit{et al.} \cite{optB}, attributing the overestimation to an excessively strong repulsive contribution in metals.

\begin{table}[H]\centering
\begin{minipage}{0.28\textwidth}
  \includegraphics[width=1.02\linewidth]{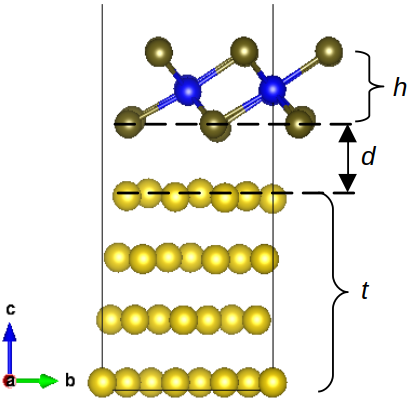}
\end{minipage}
\hspace{0.01\textwidth}
\begin{minipage}{0.68\textwidth}\centering
\begin{tabular}{llclclclc}
\textbf{System} & & \multicolumn{1}{l}{\textbf{\textit{h} (\AA)}} & & \multicolumn{1}{l}{\textbf{\textit{d} (\AA)}} &  & \multicolumn{1}{l}{\textbf{\textit{t} (\AA)}} &  & \multicolumn{1}{l}{\textbf{\textit{s} (\AA)}} \\ \hline 
\rule{0pt}{4ex}\textbf{GGA} & &  2.85  &  & 2.66  &  & 7.39  & & 2.46  \\
\rule{0pt}{4ex}\textbf{DFT-TS} & &  2.81  &  & 2.62  &  & 7.18 & & 2.39  \\
\rule{0pt}{4ex}\textbf{DFT-D3} & &  2.79  &  & 2.57 &  & 7.19 & & 2.40  \\
\rule{0pt}{4ex}\textbf{vdW-DF} & &  2.98  &  & 2.78 &  & 7.82 & & 2.61  \\
\rule{0pt}{4ex}\textbf{optB86b-vdW} & &  2.84 &  & 2.63  &  & 7.16 & & 2.39  \\
\rule{0pt}{4ex}\textbf{with SOC}  & &  2.84 &  & 2.65  &  & 7.33 & &  2.44  \\[0.2cm] \hline       
\end{tabular}
\end{minipage}
\caption{The calculated monolayer height (\textit{h}), interlayer distance (\textit{d}), and thickness (\textit{t}) and average lattice spacing (\textit{s}) of four-layer Au(111) slab obtained using different functionals for \ch{IrTe2}-Au.}\label{Table}
\end{table}